\documentclass[twocolumn,english,aps, pra, twocolumn, superscriptaddress, showpacs]{revtex4-2}
\usepackage[T1]{fontenc}
\usepackage{float}
\usepackage{amsmath}
\usepackage{amssymb}
\usepackage{graphicx}

\makeatletter

\providecommand{\tabularnewline}{\\}

\usepackage{orcidlink}

\newcommand{\UFSCar}{Departamento de Fí­sica, Universidade Federal de São Carlos, Rodovia Washington Luís, km 235 - SP-310, 13565-905 São Carlos, SP, Brasil}
\newcommand{\UFG}{Instituto de Fí­sica, Universidade Federal de Goiás, Avenida Esperança s/n, Câmpus Samambaia, 74.690-900, Goiânia - GO, Brasil}

\usepackage{hyperref}
\hypersetup{
    colorlinks=true,
    linkcolor=red,
    citecolor=blue,
    filecolor=blue,      
    urlcolor=blue,
}

\makeatother

\usepackage{babel}
\begin{document}
\title{Trapped-ion toolbox to simulate quantum Otto heat engines}
\author{Rogério Jorge de Assis~\orcidlink{0000-0003-1133-6012}}
\email{rjdeassis@gmail.com}

\affiliation{\UFSCar}
\author{Ciro Micheletti~Diniz~\orcidlink{0000-0002-7602-0468}}
\email{ciromd@outlook.com.br}

\affiliation{\UFSCar}
\author{Norton Gomes de Almeida~\orcidlink{0000-0001-8517-6774}}
\email{norton@ufg.br}

\affiliation{\UFG}
\author{Celso J. Villas Bôas~\orcidlink{0000-0001-5622-786X}}
\email{celsovb@df.ufscar.br}

\affiliation{\UFSCar}
\begin{abstract}
We present a scheme that utilizes an ion confined within a bi-dimensional
trap to simulate a quantum Otto heat engine whose working substance
is a two-level system. In this scheme, the electronic component of
the ion (the two-level system) can interact with effective heat reservoirs
of different types. We specifically focus on effective thermal reservoirs
(those with positive temperatures), effective heat reservoirs with
apparent negative temperatures, and effective squeezed thermal reservoirs.
We show how to generate these effective reservoirs and provide numerical
results to illustrate the applicability of the presented scheme. Finally,
considering the same types of effective heat reservoirs, we briefly
discuss the simulation of a quantum Otto heat engine where a quantum
harmonic oscillator serves as the working substance.
\end{abstract}
\maketitle

\section{Introduction}

In recent decades, researchers have been actively working on formulating
a theory that integrates thermodynamics with quantum mechanics, which
has come to be known as quantum thermodynamics \citep{Gemmer2009,Binder2019}.
Several studies in this area focus on investigating the so-called
quantum heat engines: heat engines in which the working substance
is a quantum system \citep{Kieu2004,Quan2007,Wang2009,Linden2010,Scully2011,Wang2012,Rahav2012,Gelbwaser-Klimovsky2013,RoBnagel2014,Uzdin2015,Beau2016,RoBnagel2016,Cakmak2017,Kosloff2017,Huang2017,Zhao2017,Klaers2017,Brandner2017,Niedenzu2018,Dorfman2018,Xiao2018,Cakmak2019,Camati2019,Turkpence2019,Peterson2019,Assis2019,Lindenfels2019,Wang2019,Wiedmann2020,Du2020,Cakmak2020,Denzler2020,Singh2020,Bouton2021,Assis2021,Miller2021,Teixeira2022,Ji2022,Kamimura2022,Kim2022,Aimet2023,Das2023}.
When studying quantum heat engines, a natural question is whether
quantum resources can improve their performance. Addressing this question,
Ref. \citep{RoBnagel2014} reveled that a quantum Otto heat engine
(QOHE) with a quantum harmonic oscillator as its working substance
can achieve higher efficiencies when, instead of a conventional thermal
reservoir (a heat reservoir with a positive temperature), its heat
source is a squeezed thermal reservoir. Furthermore, the authors showed
that the efficiency of the corresponding QOHE can surpass the Carnot
efficiency obtained from the temperatures of the thermal reservoirs
before squeezing the hottest one, which is the maximum efficiency
achievable by a heat engine operating solely with thermal reservoirs.
Following this, Ref. \citep{Klaers2017} presented an experimental
realization of such a QOHE. Employing a two-level system as the working
substance, Ref. \citep{Assis2021} demonstrated that a squeezed thermal
reservoir can increase the efficiency of the QOHE beyond the Carnot
efficiency (before squeezing) even when it operates with irreversible
unitary strokes. Here we are referring to the concept of irreversibility
from quantum thermodynamics perspective, see Ref. \citep{Plastina2014}.
Using the same working substance but now considering a heat reservoir
with an apparent negative temperature as the heat source, Ref. \citep{Assis2019}
showed that the efficiency of the QOHE can exceed the Otto efficiency
when the unitary strokes are irreversible. The potential of using
different types of heat reservoirs to improve engine efficiency is
the motivation of the present paper.

We present in this paper a scheme that employs an ion confined in
a bi-dimensional trap to simulate QOHE whose working substance is
a two-level system. The proposed scheme allows the two-level system
to operate with different kinds of effective heat reservoirs. Here
we focus on effective thermal reservoirs, effective heat reservoirs
with apparent negative temperatures, and effective squeezed thermal
reservoirs. We begin in Sec. \ref{sec:II} by describing the physical
model employed to simulate the QOHE whose working substance is the
two-level system. Then, in Secs. \ref{sec:III} and \ref{sec:IV},
we provide an overview of the unitary and non-unitary strokes that
give rise to the quantum Otto cycle. In Sec. \ref{sec:IV}, we present
the procedure for generating the effective heat reservoirs for the
two-level system. Next, in Sec. \ref{sec:V}, we describe the quantum
Otto cycle and provide the engine efficiency when it operates as a
heat engine. In Sec. \ref{sec:VI}, we show some numerical results.
Additionally, in Sec. \ref{sec:VII}, we discuss the simulation of
a QOHE with a quantum harmonic oscillator as the working substance,
considering the same types of effective heat reservoirs as above.
Finally, in Sec. \ref{sec:VIII}, we present our conclusions.

\section{\label{sec:II}Physical model}

The physical model we consider here consists of an ion within a two-dimensional
trap, allowing it to oscillate harmonically along both the $x$- and
$y$-axes. We assume here that two energy levels approximately describe
the electronic structure of the ion. Furthermore, we admit that the
ion can interact (through its dipole moment) with the electric field
of four distinct laser beams: two propagating along the $x$-direction
and two along the $y$-direction. The Hamiltonian that describes this
system at time $t$ has the form 
\begin{equation}
H^{SP}\left(t\right)=H_{e}^{SP}+H_{m}^{SP}+H_{int}^{SP}\left(t\right),\label{eq:1}
\end{equation}
with $H_{e}^{SP}$, $H_{m}^{SP}$, and $H_{int}^{SP}\left(t\right)$
being the electronic, the motional, and the interaction Hamiltonian,
respectively. The superscript ``$SP$'' indicates that these Hamiltonians
are in the Schrödinger picture. Explicitly, the Hamiltonians $H_{e}^{SP}$
and $H_{m}^{SP}$ are given by 
\begin{equation}
H_{e}^{SP}=\frac{\hbar\omega_{e}}{2}\sigma_{z}\label{eq:2}
\end{equation}
and 
\begin{equation}
H_{m}^{SP}=\sum_{\alpha=x,y}\hbar\omega_{m}a_{\alpha}^{\dagger}a_{\alpha},\label{eq:3}
\end{equation}
where $\omega_{e}$ is the electronic frequency, $\sigma_{z}$ is
the $z$-Pauli matrix, $\omega_{m}$ is the motional frequency of
the ion along the $x$- and $y$-axes (treated as equal for both axes
to simplify the model), and $a_{\alpha}$ ($a_{\alpha}^{\dagger}$)
is the respective annihilation (creation) operator, with $\alpha$
being $x$ or $y$. Considering the Lamb-Dicke regime, where the Lamb-Dicke
parameter $\lambda$ (assumed to be the same for all lasers for simplicity)
satisfies $\lambda\left[\text{tr}_{m}\bigl(a_{\alpha}^{\dagger}a_{\alpha}\bigr)\right]^{1/2}\ll1$,
the interaction Hamiltonian can be written as \citep{Leibfried2003}
\begin{multline}
H_{int}^{SP}\left(t\right)=\sum_{\alpha=x,y}\;\sum_{l=1}^{2}\frac{\hbar\Omega_{\alpha,l}}{2}\left(\sigma_{ge}+\sigma_{ge}^{\dagger}\right)\times\\
\times\left\{ 1+i\lambda\left(a_{\alpha}+a_{\alpha}^{\dagger}\right)\text{e}^{-i\left(\omega_{\alpha,l}^{L}t-\phi\right)}+\text{H.c.}\right\} ,
\end{multline}
where $\Omega_{\alpha,l}$ is the ($\alpha,l$)-Rabi frequency, $\sigma_{ge}=\left|g\right\rangle \left\langle e\right|$
($\sigma_{ge}^{\dagger}=\left|e\right\rangle \left\langle g\right|$)
is the lowering (raising) operator, with $\left|g\right\rangle $
($\left|e\right\rangle $) corresponding to the fundamental (excited)
electronic eigenstate, $\lambda$ is the Lamb-Dicke parameter, $\omega_{\alpha,l}^{L}$
is the ($\alpha,l$)-laser frequency, and $\phi$ is the initial phase
of the lasers (set to the same value for all lasers for convenience).
We assume that the dynamics of the ion state $\rho^{SP}\left(t\right)$
is that of a weak interaction with the reservoir within the Born-Markov
approximations, and therefore obeys the master equation

\begin{equation}
\dot{\rho}^{SP}\left(t\right)=\frac{1}{i\hbar}\left[H^{SP}\left(t\right),\rho^{SP}\left(t\right)\right]+\sum_{\alpha=x,y}\frac{\kappa}{2}D\left[a_{\alpha}\right]\rho^{SP}\left(t\right),
\end{equation}
in which $\kappa$ is the motional decay rate (also considered the
same for the $x$- and $y$-directions, for simplicity) and $D\left[a_{\alpha}\right]\rho^{SP}\left(t\right)=2a_{\alpha}\rho^{SP}\left(t\right)a_{\alpha}^{\dagger}-a_{\alpha}^{\dagger}a_{\alpha}\rho^{SP}\left(t\right)-\rho^{SP}\left(t\right)a_{\alpha}^{\dagger}a_{\alpha}$.
As can be seen, we neglect the electronic decay, which is a reasonable
assumption when it is much smaller than $\kappa$. This regime can
be achieved, for instance, by considering long-lived metastable electronic
states.

To simulate the QOHE described below, we switch to a rotating frame
defined by the unitary transformation 
\begin{equation}
R\left(t\right)=\text{e}^{-i\left(\omega t/2\right)\sigma_{z}}\text{e}^{-iH_{m}^{SP}t},
\end{equation}
where $\omega$ is here designated as the rotating frame frequency.
In this frame, the Hamiltonian $H^{SP}\left(t\right)$ transforms
according to $H\left(t\right)=R^{\dagger}\left(t\right)H^{SP}\left(t\right)R\left(t\right)-i\hbar R^{\dagger}\left(t\right)\dot{R}\left(t\right)$,
and can be decomposed as
\begin{equation}
H\left(t\right)=H_{e}+H_{int}\left(t\right),
\end{equation}
where 
\begin{equation}
H_{e}=\frac{\hbar\Delta_{e}}{2}\sigma_{z},
\end{equation}
with $\Delta_{e}=\omega_{e}-\omega$ being the modified electronic
frequency, and 
\begin{multline}
H_{int}\left(t\right)=\sum_{\alpha=x,y}\;\sum_{l=1}^{2}\frac{\hbar\Omega_{\alpha,l}}{2}\left(\sigma_{ge}\text{e}^{-i\omega t}+\sigma_{ge}^{\dagger}\text{e}^{i\omega t}\right)\times\\
\times\left\{ 1+i\lambda\left(a_{\alpha}\text{e}^{-i\omega_{m}t}+a_{\alpha}^{\dagger}\text{e}^{i\omega_{m}t}\right)\text{e}^{-i\left(\omega_{\alpha,l}^{L}t-\phi\right)}+\text{H.c.}\right\} .\label{eq:9}
\end{multline}
Furthermore, the ion state in the rotating frame satisfies the master
equation 
\begin{equation}
\dot{\rho}\left(t\right)=\frac{1}{i\hbar}\left[H\left(t\right),\rho\left(t\right)\right]+\sum_{\alpha=x,y}\frac{\kappa}{2}D\left[a_{\alpha}\right]\rho\left(t\right),\label{eq:10}
\end{equation}
where $\rho\left(t\right)=R^{\dagger}\left(t\right)\rho^{SP}\left(t\right)R\left(t\right)$.
All engine simulations related to the physical model described in
this section are performed within this rotating frame.

\section{\label{sec:III}Unitary stroke}

After introducing the physical model, we proceed to describe the evolution
of the two-level system throughout the strokes of the quantum Otto
cycle, beginning with the unitary strokes. In these strokes, the electronic
component of the ion (ECI) evolves unitarily while its modified frequency
changes from an initial value $\Delta_{e}\left(0\right)$ to a final
value $\Delta_{e}\left(\tau\right)$ according to a time-dependent
function $\Delta_{e}\left(t\right)$. When the modified electronic
frequency increases, the unitary stroke is called an expansion stroke;
otherwise, it is called a compression stroke.

To induce the change in the modified electronic frequency and ensure
that the ECI evolves unitarily, we consider the interaction between
the ion and an electric field of a laser beam propagating along the
$x$-direction, with a time-dependent frequency $\omega^{L}\left(t\right)\equiv\omega_{x,1}^{L}\left(t\right)=\omega\left(t\right)$.
By eliminating the summations in Eq. (\ref{eq:9}) (setting $\alpha=x$
and $l=1$) and choosing $\phi=0$, the rotating-wave approximation
yields the interaction Hamiltonian \citep{Leibfried2003} 
\begin{equation}
H_{int}=\frac{\hbar\Omega}{2}\left(\sigma_{ge}+\sigma_{ge}^{\dagger}\right),\label{eq:11}
\end{equation}
where $\Omega\equiv\Omega_{x,1}$ is the Rabi frequency. Consequently,
the dynamics of the electronic state $\rho_{e}\left(t\right)=\text{tr}_{m}\left[\rho\left(t\right)\right]$
obeys the von Neumann equation 
\begin{equation}
\dot{\rho}_{e}\left(t\right)=\frac{1}{i\hbar}\left[H_{e}\left(t\right)+H_{int},\rho_{e}\left(t\right)\right],
\end{equation}
with 
\begin{equation}
H_{e}\left(t\right)=\frac{\hbar\Delta_{e}\left(t\right)}{2}\sigma_{z},\label{eq:13}
\end{equation}
where $\Delta_{e}\left(t\right)=\omega_{e}-\omega^{L}\left(t\right)$.
Thus, the ECI evolves from the initial state $\rho_{e}\left(0\right)$
to the final state $\rho_{e}\left(\tau\right)=U_{e}\left(\tau\right)\rho_{e}\left(0\right)U_{e}^{\dagger}\left(\tau\right)$,
in which the unitary evolution operator has the form $U_{e}\left(\tau\right)=T_{+}\text{e}^{-\left(i/\hbar\right)\int_{0}^{\tau}dt\left[H_{e}\left(t\right)+H_{int}\right]}$,
with $T_{+}$ being the time-ordering operator.

By examining Eqs. (\ref{eq:11}) and (\ref{eq:13}), it is straightforward
to see that $\left[H_{e}\left(t\right)+H_{int},H_{e}\left(t'\right)+H_{int}\right]\neq0$
for any $t$ and $t'$ such that $t'\neq t$. This non-commutativity
ensures that the unitary stroke described above can induce transitions
between the electronic states $\left|g\right\rangle $ and $\left|e\right\rangle $,
depending on the duration of the stroke $\tau$. The transition probability
between these electronic states is given by 
\begin{equation}
\xi\left(\tau\right)=\left|\left\langle e\right|U_{e}\left(\tau\right)\left|g\right\rangle \right|^{2}.
\end{equation}

In most cases, transitions between $\left|g\right\rangle $ and $\left|e\right\rangle $
are undesirable because they are associated with entropy production
(irreversibility) \citep{Plastina2014}, which typically reduces engine
efficiency. For instance, this applies to the QOHE in which a two-level
system operates with two thermal reservoirs \citep{Peterson2019}
and its modified version where the hot thermal reservoir is replaced
by a squeezed one \citep{Assis2021}. In contrast, transitions during
the unitary strokes can be beneficial when the two-level system operates
with the hot heat reservoir at an apparent negative temperature, as
they may enhance engine efficiency \citep{Assis2019}. Therefore,
it is interesting to consider scenarios involving such transitions
in the unitary stroke described above.

\section{\label{sec:IV}Non-unitary stroke}

Now, we proceed to describe the non-unitary strokes, which involve
the thermalization process between the ECI and a specified effective
heat reservoir. During these strokes, we maintain the electronic frequency
constant while allowing the ion to interact, in addition to the dissipation
channels for the vibrational modes, with a specific set of laser beams,
which gives rise to the desired effective heat reservoir for its electronic
component. Below, we describe the procedure to generate effective
reservoirs such as (i) an effective thermal reservoir, (ii) an effective
heat reservoir with an apparent negative temperature, and (iii) an
effective squeezed thermal reservoir. We begin by addressing the effective
squeezed thermal reservoir, which makes it easier to approach the
remaining effective heat reservoirs.

To generate the effective squeezed thermal reservoir, we assume that
the ion interacts with the four laser beams introduced in Sec. \ref{sec:II},
see Eq. (\ref{eq:9}). Additionally, we set their frequencies as follows:
$\omega_{\alpha,1}^{L}=\omega-\omega_{m}$ and $\omega_{\alpha,2}^{L}=\omega+\omega_{m}$
($\alpha=x,y$), where $\omega$ is now a short notation for $\omega\left(0\right)$
or $\omega\left(\tau\right)$ from Sec. \ref{sec:III}. In this case,
restricting to the Lamb-Dicke regime, applying the rotating-wave approximation,
and setting $\phi=-\pi/2$, the interaction Hamiltonian (Eq. (\ref{eq:9}))
becomes 
\begin{equation}
H_{int}=\sum_{\alpha=x,y}\hbar\left(s_{\alpha}a_{\alpha}^{\dagger}+s_{\alpha}^{\dagger}a_{\alpha}\right),\label{eq:15}
\end{equation}
with 
\begin{equation}
s_{\alpha}=\frac{\lambda}{2}\left(\Omega_{\alpha,1}\sigma_{ge}+\Omega_{\alpha,2}\sigma_{ge}^{\dagger}\right).\label{eq:16}
\end{equation}
Then, taking into account Eq. (\ref{eq:15}) in Eq. (\ref{eq:10}),
assuming $\kappa\gg\lambda\Omega_{\alpha,l}$ so that $\rho\left(t\right)\approx\rho_{e}\left(t\right)\otimes\left|0_{x}\right\rangle \left\langle 0_{x}\right|\otimes\left|0_{y}\right\rangle \left\langle 0_{y}\right|$,
where $\left|0_{\alpha}\right\rangle $ is the motional ground state
in the $\alpha$ direction, and tracing over the motional degrees
of freedom, we can employ the adiabatic elimination (as detailed in
App. \ref{app:A}) \citep{Finkelstein-Shapiro2020,Saideh2020} to
obtain the effective master equation 
\begin{equation}
\dot{\rho}_{e}\left(t\right)=\frac{1}{i\hbar}\left[H_{e},\rho_{e}\left(t\right)\right]+\sum_{\alpha=x,y}\frac{2}{\kappa}D\left[s_{\alpha}\right]\rho_{e}\left(t\right),\label{eq:17}
\end{equation}
in which $H_{e}$ now represents either $H_{e}\left(0\right)$ or
$H_{e}\left(\tau\right)$, depending on the stroke considered.

Now that we have Eq. (\ref{eq:17}) at our disposal, it remains to
compare it with the master equation derived by considering a squeezed
thermal reservoir to adjust our parameters accordingly. Explicitly,
we must compare Eq. (\ref{eq:17}) with the master equation \citep{Srikanth2008}
\begin{equation}
\dot{\rho}_{e}\left(t\right)=\frac{1}{i\hbar}\left[H_{e},\rho_{e}\left(t\right)\right]+\sum_{i=1}^{2}\frac{1}{2}D\left[O_{i}\right]\rho_{e}\left(t\right),\label{eq:18}
\end{equation}
in which 
\begin{equation}
O_{1}=\sqrt{\gamma\left(1+n_{R}\right)}\left(\mu\sigma_{ge}+\nu\sigma_{ge}^{\dagger}\right),
\end{equation}
and 
\begin{equation}
O_{2}=\sqrt{\gamma n_{R}}\left(\nu\sigma_{ge}+\mu\sigma_{ge}^{\dagger}\right).
\end{equation}
Here, $\gamma$ is the electronic decay rate (related to the coupling
of the ion to a squeezed thermal reservoir), $n_{R}$ is the average
number of photons in the thermal reservoir before squeezing (which
follows from the Bose-Einstein distribution), $\mu=\cosh\left(r\right)$,
and $\nu=\sinh\left(r\right)$, with $r$ being the squeezing parameter.
Upon comparing Eq. (\ref{eq:17}) with Eq. (\ref{eq:18}), it becomes
evident that if we chose 
\begin{equation}
\frac{\lambda\Omega_{x,1}}{\sqrt{\kappa}}=\sqrt{\gamma\left(1+n_{R}\right)}\mu,\label{eq:21}
\end{equation}
\begin{equation}
\frac{\lambda\Omega_{x,2}}{\sqrt{\kappa}}=\sqrt{\gamma\left(1+n_{R}\right)}\nu,
\end{equation}
\begin{equation}
\frac{\lambda\Omega_{y,1}}{\sqrt{\kappa}}=\sqrt{\gamma n_{R}}\nu,
\end{equation}
and 
\begin{equation}
\frac{\lambda\Omega_{y,2}}{\sqrt{\kappa}}=\sqrt{\gamma n_{R}}\mu,\label{eq:24}
\end{equation}
we ensure that Eq. (\ref{eq:17}) reproduces the dynamics of a two-level
system weakly coupled to a squeezed thermal reservoir.

Before proceeding to the remaining effective heat reservoirs, it is
worthwhile to show the state that the ECI reaches when equilibrating
with the effective squeezed thermal reservoir. By solving Eq. (\ref{eq:18})
and taking $t\rightarrow\infty$, we obtain the steady-state \citep{Srikanth2008}
\begin{equation}
\rho_{e}^{S}=S_{\mu,\nu}\left(\rho_{e}^{G}\right),\label{eq:25}
\end{equation}
in which $\rho_{e}^{G}$ is the so-called Gibbs state, and $S_{\mu,\nu}\left(.\right)$
is such that 
\begin{equation}
S_{\mu,\nu}\left(\left|g\right\rangle \left\langle g\right|\right)=\frac{1}{\mu^{2}+\nu^{2}}\left(\mu^{2}\left|g\right\rangle \left\langle g\right|+\nu^{2}\left|e\right\rangle \left\langle e\right|\right),\label{eq:26}
\end{equation}
and 
\begin{gather}
S_{\mu,\nu}\left(\left|e\right\rangle \left\langle e\right|\right)=\frac{1}{\mu^{2}+\nu^{2}}\left(\nu^{2}\left|g\right\rangle \left\langle g\right|+\mu^{2}\left|e\right\rangle \left\langle e\right|\right).\label{eq:27}
\end{gather}
The Gibbs state has the form 
\begin{equation}
\rho_{e}^{G}=\frac{\text{e}^{-\beta_{R}H_{e}}}{Z_{e}},\label{eq:28}
\end{equation}
where $\beta_{R}=1/\left(k_{B}T_{R}\right)$, with $T_{R}$ being
the temperature of the effective thermal reservoir before squeezing,
and $Z_{e}=\text{tr}\left(\text{e}^{-\beta_{R}H_{e}}\right)$. So,
if Eqs. (\ref{eq:21})-(\ref{eq:24}) are satisfied, $\rho_{e}^{S}$
is the electronic state after equilibration.

We now address the effective thermal reservoir either at positive
or apparent negative temperatures. The procedure we follow here is
similar to the one performed above: we compare Eq. (\ref{eq:17})
with the master equation obtained by considering the desired heat
reservoir, aiming to identify the conditions that make these equations
identical. The master equation we must compare with Eq. (\ref{eq:17})
follows the same structure as Eq. (\ref{eq:18}) but with operators
\citep{Li2011,Ghosh2012} 
\begin{equation}
O_{1}=\sqrt{\gamma\left(1\pm n_{R}\right)}\sigma_{ge}\label{eq:29}
\end{equation}
and 
\begin{equation}
O_{2}=\sqrt{\gamma n_{R}}\sigma_{ge}^{\dagger}.\label{eq:30}
\end{equation}
In Eq. (\ref{eq:30}), we apply the positive sign if $n_{R}=1/\left(\text{e}^{\beta_{R}\hbar\Delta_{e}}-1\right)$,
following the Bose-Einstein distribution ($n_{R}\in\left(0,\infty\right)$),
and the negative sign if $n_{R}=1/\left(\text{e}^{\beta_{R}\hbar\Delta_{e}}+1\right)$,
following the Fermi-Dirac distribution ($n_{R}\in\left(0,1\right)$).
When $n_{R}$ corresponds to the Fermi-Dirac distribution and $n_{R}\in\left(1/2,1\right)$,
$\beta_{R}<0$; for $n_{R}\in\left(0,1/2\right)$, $\beta_{R}>0$.
Therefore, when we consider $O_{1}$ and $O_{2}$ as defined by Eqs.
(\ref{eq:29}) and (\ref{eq:30}), Eq. (\ref{eq:18}) covers both
master equations corresponding to a thermal reservoir ($\beta_{R}>0$)
and a heat reservoir with an apparent negative temperature ($\beta_{R}<0$).
Then, by comparing this equation with Eq. (\ref{eq:17}), we obtain
\begin{equation}
\frac{\lambda\Omega_{x,1}}{\sqrt{\kappa}}=\sqrt{\gamma\left(1\pm n_{R}\right)},\label{eq:31}
\end{equation}
\begin{equation}
\frac{\lambda\Omega_{x,2}}{\sqrt{\kappa}}=0,
\end{equation}
\begin{equation}
\frac{\lambda\Omega_{y,1}}{\sqrt{\kappa}}=0,
\end{equation}
and 
\begin{equation}
\frac{\lambda\Omega_{y,2}}{\sqrt{\kappa}}=\sqrt{\gamma n_{R}}.\label{eq:34}
\end{equation}
Thus, if we configure the laser beams to satisfy Eqs. (\ref{eq:31})-(\ref{eq:34}),
Eq. (\ref{eq:17}) reproduces the dynamics of a two-level system in
contact with a thermal reservoir, either at positive or at apparent
negative temperatures. As it is well known, under the conditions above
the steady-state of Eq. (\ref{eq:17}) is the Gibbs state displayed
in Eq. (\ref{eq:28}), with $\beta_{R}>0$ or $\beta_{R}<0$, depending
on $O_{1}$ and $O_{2}$.

\begin{figure*}[t]
\begin{centering}
\includegraphics{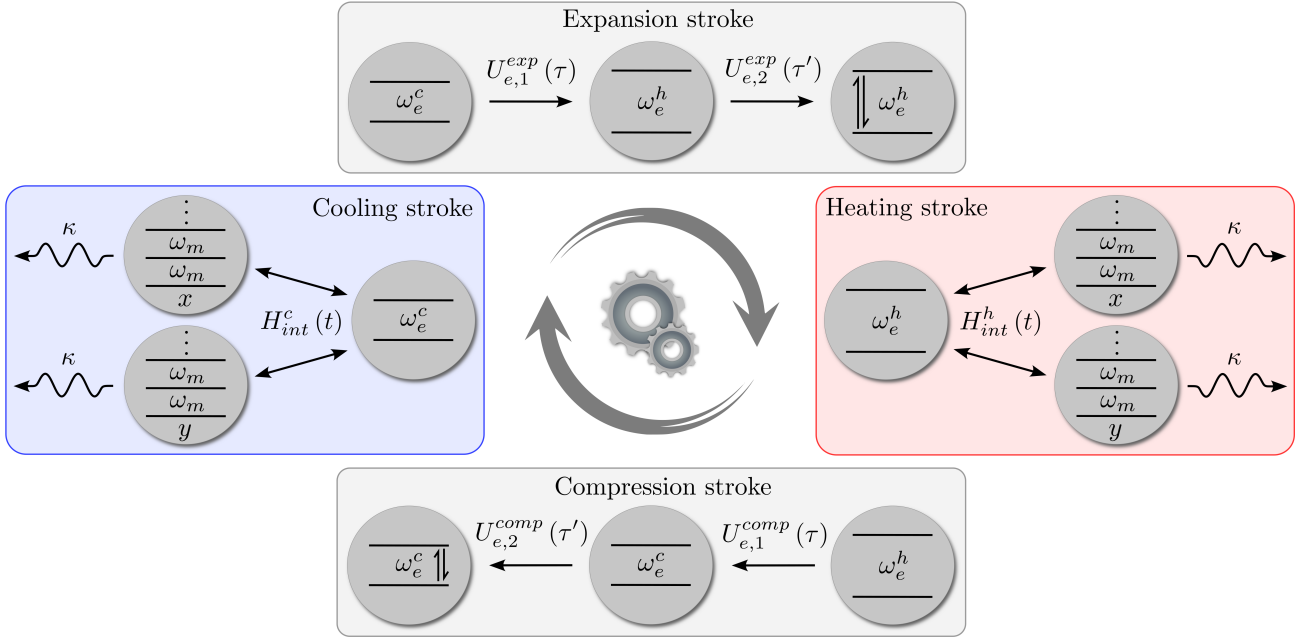}
\par\end{centering}
\caption{\label{fig:1}Scheme displaying the quantum Otto cycle described in
the text. The boxes represent the cycle strokes. Starting with the
cooling stroke, $\Delta_{e}^{c}$ corresponds to the modified electronic
frequency of the ion, $\omega_{m}$ represents the motional frequency
of the ion along the $x$- and $y$-directions, $\kappa$ denotes
the motional decay rate, and $H_{int}^{c}$ stands for the interaction
Hamiltonian between the electronic and motional components of the
ion. As the text explains, $H_{int}^{c}$ determines the effective
cold heat reservoir for the electronic component of the ion (ECI).
In the expansion stroke, the two-level system undergoes a unitary
evolution governed by $U_{e}^{exp}\left(\tau\right)$, which changes
the modified electronic frequency from $\Delta_{e}^{c}$ to $\Delta_{e}^{h}$
($\Delta_{e}^{c}<\Delta_{e}^{h}$) after time $\tau$. During the
heating stroke, the interaction Hamiltonian is now $H_{int}^{h}$,
which leads to the effective hot heat reservoir for the ECI. Finally,
in the compression stroke, the unitary operator $U_{e}^{comp}\left(\tau\right)$
reverses the change, bringing the modified electronic frequency back
from from $\Delta_{e}^{h}$ to $\Delta_{e}^{c}$.}
\end{figure*}

\section{\label{sec:V}Quantum Otto heat engine}

We begin this section by describing the quantum Otto cycle corresponding
to the QOHE we are interested in simulating. As previously mentioned,
our goal here is to simulate in the context of trapped ions a QOHE
whose working substance is a two-level system operating between a
cold and a hot heat reservoir. We assume that the cold heat reservoir
is always a thermal reservoir. On the other hand, the hot heat reservoir
can be a thermal reservoir, a heat reservoir with an apparent negative
temperature, or a squeezed thermal reservoir. As discussed previously,
the two-level system corresponds to the ECI, which can effectively
interact with these reservoirs. To facilitate the exposition, we initially
describe the quantum Otto cycle with an effective squeezed thermal
reservoir as its hot heat reservoir. The strokes that constitute the
quantum Otto cycle are as follows (see Fig. \ref{fig:1}):

\medskip{}

\noindent \textbf{Expansion stroke.} In this first stroke, the ECI
undergoes the unitary evolution described by $U_{e}^{exp}\left(\tau\right)$,
which corresponds to the unitary operator $U_{e}\left(\tau\right)$
introduced in Sec. \ref{sec:III}, with $\Delta_{e}\left(0\right)=\Delta_{e}^{c}$
and $\Delta_{e}\left(\tau\right)=\Delta_{e}^{h}$ ($\Delta_{e}^{c}<\Delta_{e}^{h}$).
Here, we consider that $\Delta_{e}^{c}$ switches to $\Delta_{e}^{h}$
according to $\Delta_{e}\left(t\right)=\left(1-t/\tau\right)\Delta_{e}^{c}+\left(t/\tau\right)\Delta_{e}^{h}$.
At the beginning of this stroke, the ECI is in the Gibbs state $\rho_{e}^{G,c}=\text{e}^{-\beta_{R}^{c}H_{e}^{c}}/Z_{e}^{c}$,
in which $\beta_{R}^{c}=1/k_{B}T_{R}^{c}$, with $T_{R}^{c}$ being
the temperature of the effective cold thermal reservoir, $H_{e}^{c}=\hbar\Delta_{e}^{c}\sigma_{z}/2$,
and $Z_{e}^{c}=\text{tr}\bigl(\text{e}^{-\beta_{R}^{c}H_{e}^{c}}\bigr)$.
Note that the superscript ``$c$'' denotes that we are referring
to the effective cold heat reservoir. Then, the electronic state at
the end of the unitary evolution is $\rho_{e}^{exp}\left(\tau\right)=U_{e}^{exp}\left(\tau\right)\rho_{e}^{G,c}\;U_{e}^{exp,\dagger}\left(\tau\right)$.
\medskip{}

\noindent \textbf{Heating stroke.} At this stroke, the ECI evolves
non-unitarily by interacting effectively with the squeezed hot thermal
reservoir, as addressed in Sec. \ref{sec:IV}. Therefore, the dynamics
of the electronic state is governed by Eq. (\ref{eq:17}), with the
parameters meeting the conditions specified in Eqs. (\ref{eq:21})-(\ref{eq:24}).
To explicitly indicate that we are dealing with the effective hot
heat reservoir, we use the superscript ``$h$'' in these equations,
resulting in the notation changes $H_{e}\rightarrow H_{e}^{h}=\hbar\Delta_{e}^{h}\sigma_{z}/2$,
$s_{\alpha}\rightarrow s_{\alpha}^{h}=\lambda\bigl(\Omega_{\alpha,1}^{h}\sigma_{-}+\Omega_{\alpha,2}^{h}\sigma_{+}\bigr)/2$,
and $n_{R}\rightarrow n_{R}^{h}$. The stroke ends when the ECI equilibrates
with the effective hot heat reservoir, reaching the steady-state $\rho_{e}^{S,h}=S_{\mu,\nu}\left(\rho_{e}^{G,h}\right)$,
where $\rho_{e}^{G,h}=\text{e}^{-\beta_{R}^{h}H_{e}^{h}}/Z_{e}^{h}$.
Here, $\beta_{R}^{h}=1/k_{B}T_{R}^{h}$, with $T_{R}^{h}$ being the
temperature of the hot thermal reservoir before squeezing, and $Z_{e}^{h}=\text{tr}\bigl(\text{e}^{-\beta_{R}^{h}H_{e}^{h}}\bigr)$.
\medskip{}

\noindent \textbf{Compression stroke.} During this stroke, the electronic
state evolves unitarily according to $U_{e}^{comp}\left(\tau\right)=U_{e}^{exp,\dagger}\left(\tau\right)$,
which matches the adjoint of $U_{e}\left(\tau\right)$ from Sec. \ref{sec:III},
observing that now $\Delta_{e}\left(0\right)=\Delta_{e}^{h}$ and
$\Delta_{e}\left(\tau\right)=\Delta_{e}^{c}$ (here, $\Delta_{e}\left(t\right)=\left(1-t/\tau\right)\Delta_{e}^{h}+\left(t/\tau\right)\Delta_{e}^{c}$).
As a result, the present stroke leads the ECI to the state $\rho_{e}^{comp}\left(\tau\right)=U_{e}^{comp}\left(\tau\right)\rho_{e}^{S,h}\;U_{e}^{comp,\dagger}\left(\tau\right)$.\medskip{}

\noindent \textbf{Cooling stroke.} As a final stroke, the ECI interacts
effectively with the cold thermal reservoir until thermalization.
Here, as discussed in Sec. \ref{sec:IV}, the electronic state undergoes
a non-unitary evolution dictated by Eq. (\ref{eq:17}), obeying Eqs.
(\ref{eq:31})-(\ref{eq:34}). To explicitly denote that we are referring
to the effective cold heat reservoir, we introduce the superscript
``$c$'' in these equations, leading to the notation modifications
$H_{e}\rightarrow H_{e}^{c}$, $s_{\alpha}\rightarrow s_{\alpha}^{c}=\lambda\bigl(\Omega_{\alpha,1}^{c}\sigma_{-}+\Omega_{\alpha,2}^{c}\sigma_{+}\bigr)/2$,
and $n_{R}\rightarrow n_{R}^{c}$. When the thermalization occurs,
the ECI reaches the Gibbs state $\rho_{e}^{G,c}$, thus closing the
cycle.\medskip{}

After describing the cycle, we can proceed to determine the engine
efficiency. For this purpose, we first calculate the heat and work
associated with each cycle stroke. To obtain these quantities, we
adopt the following definitions for work and heat \citep{Alicki1979,Kosloff1984}:
$W_{e}^{str}=\int_{t_{i}}^{t_{f}}dt\text{tr}\bigl[\dot{H}_{e}^{str}\left(t\right)\rho_{e}^{str}\left(t\right)\bigr]$
and $Q_{e}^{str}=\int_{t_{i}}^{t_{f}}dt\text{tr}\left[H_{e}^{str}\left(t\right)\dot{\rho}_{e}^{str}\left(t\right)\right]$.
Here, $H_{e}^{str}\left(t\right)$ and $\rho_{e}^{str}\left(t\right)$
correspond to the Hamiltonian and the state that describe the ECI
throughout the respective strokes. With these definitions, it is straightforward
to show that the expansion and compression strokes involve only work
($W_{e}^{comp}$ and $W_{e}^{exp}$), whereas the heating and cooling
strokes involve solely heat ($Q_{e}^{h}$ and $Q_{e}^{c}$). Thus,
by calculating the energy exchanged in each stroke, we obtain the
net work ($W_{e}^{net}=W_{e}^{exp}+W_{e}^{comp}$) 
\begin{multline}
W_{e}^{net}=-\frac{\hbar\left(\Delta_{e}^{h}-\Delta_{e}^{c}\right)}{2}\left[\tanh\left(\theta^{c}\right)-\zeta\tanh\left(\theta^{h}\right)\right]\\
+\hbar\xi\left[\Delta_{e}^{h}\tanh\left(\theta^{c}\right)+\Delta_{e}^{c}\zeta\tanh\left(\theta^{h}\right)\right],\label{eq:35}
\end{multline}
the heat associated with the heating stroke 
\begin{equation}
Q_{e}^{h}=\frac{\hbar\Delta_{e}^{h}}{2}\left[\tanh\left(\theta^{c}\right)-\zeta\tanh\left(\theta^{h}\right)-2\xi\tanh\left(\theta^{c}\right)\right],
\end{equation}
and the heat linked to the cooling stroke 
\begin{equation}
Q_{e}^{c}=-\frac{\hbar\Delta_{e}^{c}}{2}\left[\tanh\left(\theta^{c}\right)-\zeta\tanh\left(\theta^{h}\right)+2\xi\zeta\tanh\left(\theta^{h}\right)\right],\label{eq:37}
\end{equation}
where $\theta^{c\left(h\right)}=\beta_{R}^{c\left(h\right)}\hbar\Delta_{e}^{c\left(h\right)}/2$
and $\zeta=1/\left(\mu^{2}+\nu^{2}\right)$, as derived in Ref. \citep{Assis2021}.
For the sake of simplicity, we are omitting the time dependence of
the probability transition ($\xi\equiv\xi\left(\tau\right)$).

\begin{figure*}[t]
\begin{centering}
\includegraphics{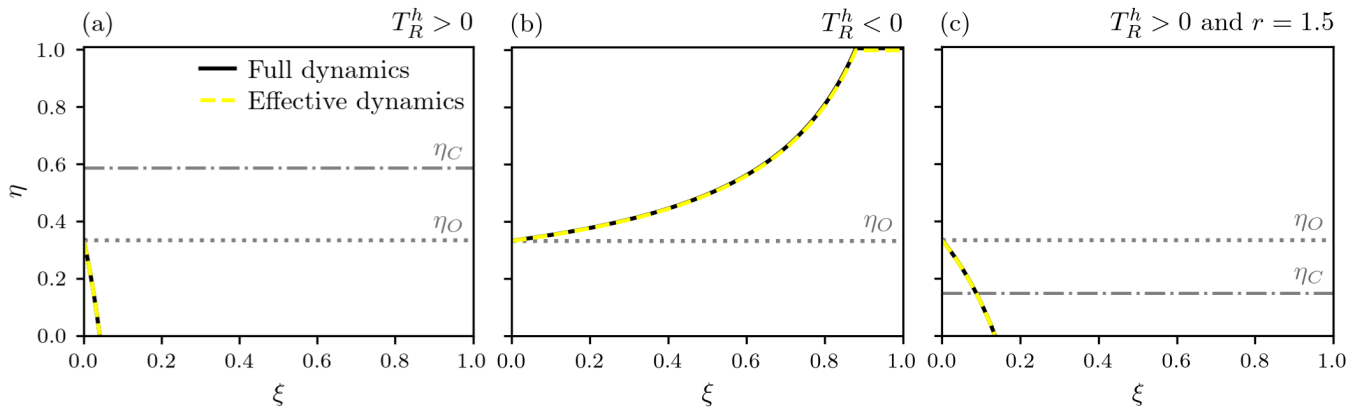}
\par\end{centering}
\caption{\label{fig:2}Engine efficiency $\eta$ as a function of the transition
probability $\xi$. Panel (a) displays curves associated with the
effective hot thermal reservoir, panel (b) shows curves related to
the effective hot reservoir with an apparent negative temperature,
and panel (c) depicts curves linked with the effective hot squeezed
thermal reservoir. In panels (a)-(c), we assume a cold thermal reservoir
with $n_{R}^{c}=0.6$, defined by the Bose-Einstein distribution.
In panel (a), we choose $n_{R}^{h}=1.2$, determined by the Bose-Einstein
distribution; in panel (b), $n_{R}^{h}=0.8$, given by the Fermi-Dirac
distribution; and in panel (c), $n_{R}^{h}=0.4$, provided by the
Bose-Einstein distribution, and $r=1.5$. In addition, we set $\Delta_{e}^{c}=2\pi\times\left(0.2\ \text{MHz}\right)$,
$\Delta_{e}^{h}=2\pi\times\left(0.4\ \text{MHz}\right)$, $\lambda=10^{-2}$,
$\Omega=2\pi\times\left(10\ \text{MHz}\right)$, $\kappa=2\pi\times\left(1\ \text{MHz}\right)$,
and $\gamma=2\pi\times\left(10^{-3}\ \text{MHz}\right)$. The red
dots correspond to the case in which we apply the adiabatic elimination
in the heating and cooling strokes (effective dynamics), the black
solid lines correspond to the situation in which we do not perform
the adiabatic elimination (full dynamics), the gray dotted lines show
the Otto efficiency $\eta_{O}$, and the gray dot-dashed lines display
the Carnot efficiency $\eta_{C}$.}
\end{figure*}

The condition for the cycle presented above to operate as a heat engine
is $W_{e}^{net}<0$, indicating the extraction of net work from the
ECI after the four strokes. If this condition is satisfied, the engine
efficiency $\eta$ is determined using the formula 
\begin{equation}
\eta=-\frac{W_{e}^{net}}{Q_{e}^{abs}},\label{eq:38}
\end{equation}
with $Q_{e}^{abs}$ representing the heat absorbed by the ECI ($Q_{e}^{abs}>0$)
during the cycle. Considering Eqs. (\ref{eq:35})-(\ref{eq:37}),
it is easy to demonstrate that $W_{e}^{net}<0$ implies $Q_{e}^{h}>0$
and $Q_{e}^{c}<0$, resulting in $Q_{e}^{abs}=Q_{e}^{h}$. As a result,
the engine efficiency can be expressed as \citep{Assis2021}
\begin{equation}
\eta^{S}=1-\frac{\Delta_{e}^{c}}{\Delta_{e}^{h}}\left[\frac{\tanh\left(\theta^{c}\right)-\left(1-2\xi\right)\zeta\tanh\left(\theta^{h}\right)}{\left(1-2\xi\right)\tanh\left(\theta^{c}\right)-\zeta\tanh\left(\theta^{h}\right)}\right].
\end{equation}
As stated in Ref. \citep{Assis2021}, $\eta^{S}$ cannot surpass the
Otto efficiency $\eta_{O}=1-\Delta_{e}^{c}/\Delta_{e}^{h}$ but can
exceed the Carnot efficiency $\eta_{C}=1-\beta_{R}^{h}/\beta_{R}^{c}$.
This is because the squeezing process actually adds photons into the
hot heat reservoir, increasing its effective temperature when compared
to its initial temperature $T_{R}^{h}$.

We can now easily adjust the cycle and expressions introduced above
to account for the cases where the QOHE operates with a thermal reservoir
and a heat reservoir with an apparent negative temperature as its
hot heat reservoir. The adjustment in the cycle involves allowing
the ECI to interact effectively with one of these heat reservoirs,
which changes the final electronic state of the heating stroke and,
consequently, the initial and final electronic states of the compression
stroke. Thus, the ECI attains a Gibbs state (Eq. (\ref{eq:28})) at
the end of the heating stroke, with $\beta_{R}^{h}>0$ for the effective
hot thermal reservoir and $\beta_{R}^{h}<0$ for the effective hot
heat reservoir with an apparent negative temperature. Subsequently,
we obtain the final electronic state of the compression stroke by
applying $U_{e}^{comp}\left(\tau\right)$ to the respective Gibbs
state. As a result, we can rewrite Eqs. (\ref{eq:35})-(\ref{eq:37})
by taking $r=0$ (which implies $S_{\mu,\nu}\left(.\right)=I$ and
$\zeta=1$) and, if we are considering an effective hot heat reservoir
with an apparent negative temperature, $\beta_{R}^{h}=-\left|\beta_{R}^{h}\right|$
($\theta^{h}=-\left|\theta^{h}\right|$). In the case involving an
effective hot thermal reservoir, $W_{e}^{net}<0$ then results in
$Q_{e}^{h}>0$ and $Q_{e}^{c}<0$ ($Q_{e}^{abs}=Q_{e}^{h}$), allowing
us to express the engine efficiency as \citep{Peterson2019} 
\begin{equation}
\eta^{+}=1-\frac{\Delta_{e}^{c}}{\Delta_{e}^{h}}\left[\frac{\tanh\left(\theta^{c}\right)-\left(1-2\xi\right)\tanh\left(\theta^{h}\right)}{\left(1-2\xi\right)\tanh\left(\theta^{c}\right)-\tanh\left(\theta^{h}\right)}\right].
\end{equation}
On the other hand, in the case involving an effective hot heat reservoir
with an apparent negative temperature, $W_{e}^{net}<0$ does not always
imply $Q_{e}^{h}>0$ and $Q_{e}^{c}<0$; there are values of $\xi$
that yield $Q_{e}^{h}>0$ and $Q_{e}^{c}>0$. If the ECI absorbs heat
from both effective heat reservoirs, $\eta^{-}=1$, whereas if it
absorbs heat only from the effective hot heat reservoir \citep{Assis2019},
\begin{equation}
\eta^{-}=1-\frac{\Delta_{e}^{c}}{\Delta_{e}^{h}}\left[\frac{\tanh\left(\theta^{c}\right)+\left(1-2\xi\right)\tanh\left(\left|\theta^{h}\right|\right)}{\left(1-2\xi\right)\tanh\left(\theta^{c}\right)+\tanh\left(\left|\theta^{h}\right|\right)}\right].\label{eq:41}
\end{equation}
As discussed in Ref. \citep{Assis2019}, the hot heat reservoir with
an apparent negative temperature can lead to $\eta^{-}>\eta_{O}$,
depending on the value of $\xi$.

\section{\label{sec:VI}Numerical results}

In this section, we present some numerical results that illustrate
the effectiveness of the proposed scheme in simulating the QOHE of
interest. For this purpose, we numerically replicate the cycles described
in the previous section, considering both the scenario in which we
perform the adiabatic elimination (in the heating and cooling strokes)
and the one where we do not. By applying this approximation, we obtain
the dynamics of the ECI utilizing Eq. (\ref{eq:17}); otherwise, we
use Eq. (\ref{eq:10}) with the interaction Hamiltonian provided by
Eq. (\ref{eq:15}). In this way, we can verify the accuracy of the
adiabatic elimination, which gives rise to the desired effective dynamics
during the heating and cooling strokes. Furthermore, while ensuring
the Lamb-Dicke regime and the validity of the rotating wave approximation,
we select parameters of the same order as those observed in experimental
setups involving trapped ions \citep{Leibfried2003,Haffner2008,Blatt2012}.
We obtain the numerical results using the Quantum Toolbox in Python
(QuTiP) \citep{Johansson2012,Johansson2013}.

The numerical results are shown in Figs. \ref{fig:2}(a)-(c), where
we display the engine efficiency $\eta$ as a function of the transition
probability $\xi$. Fig. \ref{fig:2}(a) corresponds to the case involving
an effective hot thermal reservoir ($\eta^{+}$), Fig. \ref{fig:2}(b)
to the situation involving an effective hot heat reservoir with an
apparent negative temperature ($\eta^{-}$), and Fig. \ref{fig:2}(c)
to the scenario involving an effective squeezed hot thermal reservoir
($\eta^{S}$). In these figures, we set $\Delta_{e}^{c}=2\pi\times\left(0.2\ \text{MHz}\right)$,
$\Delta_{e}^{h}=2\pi\times\left(0.4\ \text{MHz}\right)$, $\lambda=10^{-2}$,
$\Omega=2\pi\times\left(10\ \text{MHz}\right)$, $\kappa=2\pi\times\left(1\ \text{MHz}\right)$,
$\gamma=2\pi\times\left(10^{-3}\ \text{MHz}\right)$, and $n_{R}^{c}=0.6$
(given by the Bose-Einstein distribution). Additionally, we assume
$n_{R}^{h}=1.2$ (provided by the Bose-Einstein distribution) in Fig.
\ref{fig:2}(a), $n_{R}^{h}=0.8$ (given by the Fermi-Dirac distribution)
in Fig. \ref{fig:2}(b), and $n_{R}^{h}=0.4$ (determined by the Bose-Einstein
distribution) and $r=1.5$ in Fig. \ref{fig:2}(c). With these quantities,
it is straightforward to calculate the respective Rabi frequencies
$\Omega_{x,1}^{c\left(h\right)}$, $\Omega_{x,2}^{c\left(h\right)}$,
$\Omega_{y,1}^{c\left(h\right)}$, and $\Omega_{y,2}^{c\left(h\right)}$
by using Eqs. (\ref{eq:21})-(\ref{eq:24}) or (\ref{eq:31})-(\ref{eq:34}),
depending on the case. Therefore, since the parameters are adjusted
to reproduce the desired dynamics, it remains solely to verify the
validity of the approximations employed to obtain the effective dynamics.
In Figs. \ref{fig:2}(a)-(c), the red dots correspond to the case
in which we perform the adiabatic elimination (effective dynamics),
derived in detail in App. \ref{app:A}, while the solid black lines
are associated with the scenario in which we do not perform such an
approximation (full dynamics). Besides, the gray dotted and dot-dashed
lines indicate the corresponding Otto and Carnot efficiencies, respectively.
As we can see, the red dots overlap the black solid ones, thus illustrating
the validity of the adiabatic elimination applied to derive Eq. (\ref{eq:18}).

\section{\label{sec:VII}How about a quantum harmonic oscillator as the working
substance?}

Having addressed the case involving a two-level system as the working
substance, we now briefly discuss the simulation of a QOHE that has
a quantum harmonic oscillator as the working substance. To simulate
such a QOHE, considering the different types of heat reservoirs discussed
previously, we need a physical model different from the one presented
in Sec. \ref{sec:II}. Here, considering only the oscillation of the
ion in a given direction, assuming that three energy levels in the
V configuration define its electronic structure, and allowing it to
interact with a set of four laser beams (as in Sec. \ref{sec:II}),
the Hamiltonian that describes the ion is 
\begin{equation}
H^{SP}\left(t\right)=H_{e}^{SP}+H_{m}^{SP}+H^{SP}\left(t\right),
\end{equation}
with 
\begin{equation}
H_{e}^{SP}=\frac{\hbar\omega_{ge}}{2}\left|e\right\rangle \left\langle e\right|+\frac{\hbar\omega_{gf}}{2}\left|f\right\rangle \left\langle f\right|,\label{eq:43}
\end{equation}
\begin{equation}
H_{m}^{SP}=\hbar\omega_{m}a^{\dagger}a,
\end{equation}
and (in the Lamb-Dicke regime)
\begin{multline}
H_{int}^{SP}\left(t\right)=\sum_{\alpha=ge,gf}\;\sum_{l=1}^{2}\frac{\hbar\Omega_{\alpha,l}}{2}\left(\sigma_{\alpha}+\sigma_{\alpha}^{\dagger}\right)\times\\
\times\left\{ 1+i\lambda\left(a+a^{\dagger}\right)\text{e}^{-i\left(\omega_{\alpha,l}^{L}t-\phi\right)}+\text{H.c.}\right\} .\label{eq:45}
\end{multline}
In Eq. (\ref{eq:43}), $\left|e\right\rangle $ and $\left|f\right\rangle $
are the excited electronic states, while $\omega_{ge}$ and $\omega_{gf}$
are the electronic transition frequencies between them and the ground
state $\left|g\right\rangle $; in Eq. (\ref{eq:45}), $\sigma_{gf}=\left|g\right\rangle \left\langle f\right|$
($\sigma_{gf}^{\dagger}=\left|f\right\rangle \left\langle g\right|$).
Furthermore, we assume that the dynamics of the ion state is given
now by the master equation 
\begin{equation}
\dot{\rho}^{SP}\left(t\right)=\frac{1}{i\hbar}\left[H^{SP}\left(t\right),\rho^{SP}\left(t\right)\right]+\sum_{\alpha=ge,gf}\frac{\gamma_{\alpha}}{2}D\left[\sigma_{\alpha}\right]\rho^{SP}\left(t\right),
\end{equation}
where $\gamma_{ge}$ and $\gamma_{gf}$ are the electronic decay rates.
Here, in contrast to Sec. \ref{sec:II}, we consider the motional
decay rate to be significantly smaller than the electronic decay rates,
which can be achieved, for instance, by working with motional modes
that inherently possess long coherence times. Consequently, we can
safely disregard it.

Now, similarly to Sec. \ref{sec:II}, we switch to the rotating frame
defined by the unitary transformation
\begin{equation}
R\left(t\right)=\text{e}^{-iH_{e}^{SP}t}\text{e}^{-i\left(\omega t\right)a^{\dagger}a},
\end{equation}
resulting in the master equation
\begin{equation}
\dot{\rho}\left(t\right)=\frac{1}{i\hbar}\left[H\left(t\right),\rho\left(t\right)\right]+\sum_{\alpha=ge,gf}\frac{\gamma_{\alpha}}{2}D\left[\sigma_{\alpha}\right]\rho\left(t\right),\label{eq:48}
\end{equation}
in which
\begin{equation}
H\left(t\right)=H_{m}+H_{int}\left(t\right),
\end{equation}
with
\begin{equation}
H_{m}=\hbar\Delta_{m}a^{\dagger}a,
\end{equation}
being $\Delta_{m}=\omega_{m}-\omega$, and
\begin{multline}
H_{int}\left(t\right)=\sum_{\alpha=ge,gf}\;\sum_{l=1}^{2}\frac{\hbar\Omega_{\alpha,l}}{2}\left(\sigma_{\alpha}\text{e}^{-i\omega_{\alpha}t}+\sigma_{\alpha}^{\dagger}\text{e}^{i\omega_{\alpha}t}\right)\times\\
\times\left\{ 1+i\lambda\left(a\text{e}^{-i\omega t}+a^{\dagger}\text{e}^{i\omega t}\right)\text{e}^{-i\left(\omega_{\alpha,l}^{L}t-\phi\right)}+\text{H.c.}\right\} .\label{eq:51}
\end{multline}

Having defined the physical model, we can now describe the generation
of the effective heat reservoirs of interest for the quantum harmonic
oscillator. Following the same strategy as in Sec. \ref{sec:IV},
we start by setting $\omega_{\alpha,1}^{L}=\omega-\omega_{\alpha}$,
$\omega_{\alpha,2}^{L}=\omega+\omega_{\alpha}$, and $\phi=-\pi/2$.
As a result, the rotating-wave approximation lead to the interaction
Hamiltonian 
\begin{equation}
H_{int}=\sum_{\alpha=ge,gf}\hbar\left(s_{\alpha}\sigma_{\alpha}^{\dagger}+s_{\alpha}^{\dagger}\sigma_{\alpha}\right),\label{eq:52}
\end{equation}
in which 
\begin{equation}
s_{\alpha}=\frac{\lambda}{2}\left(\Omega_{\alpha,1}a+\Omega_{\alpha,2}a^{\dagger}\right).
\end{equation}
Then, assuming $\gamma_{\alpha}\gg\lambda\Omega_{\alpha,l}$ so that
$\rho\left(t\right)\approx\left|g\right\rangle \left\langle g\right|\otimes\rho_{m}\left(t\right)$,
tracing over the electronic degree of freedom, and applying the adiabatic
elimination (see App. \ref{app:B}), we obtain the effective master
equation 
\begin{equation}
\dot{\rho}_{m}\left(t\right)=\frac{1}{i\hbar}\left[H_{m},\rho_{m}\left(t\right)\right]+\sum_{\alpha=ge,gf}\frac{2}{\gamma_{\alpha}}D\left[s_{\alpha}\right]\rho_{m}\left(t\right).\label{eq:54}
\end{equation}
Thus, when we properly choose the Rabi frequencies, obeying similar
relations to the ones of Eqs. (\ref{eq:21})-(\ref{eq:24}) and (\ref{eq:31})-(\ref{eq:34}),
Eq. (\ref{eq:54}) can effectively describe the dynamics of the quantum
harmonic oscillator in contact with the heat reservoirs discussed
in the previous sections.

Finally, we implement the unitary steps of the cycle by turning off
the laser beams and modifying the trap potential, which changes $\omega_{m}$
and, consequently, $\Delta_{m}$. Note that variations in $\omega_{m}$
can induce transitions in the motional part of the ion, since $\left[H_{m}\left(t\right),H_{m}\left(t'\right)\right]\neq0$
for any $t$ and $t'$ such that $t'\neq t$.

\section{\label{sec:VIII}Conclusion}

We present a scheme that utilizes an ion confined within a bi-dimensional
trap to simulate a quantum Otto heat engine (QOHE) with a two-level
system as its working substance. Our scheme allows the electronic
component of the ion (ECI), here considered as a two-level system,
to interact with different types of effective heat reservoirs, including
effective thermal reservoirs (effective heat reservoirs with positive
temperatures), effective heat reservoirs with apparent negative temperatures,
and effective squeezed thermal reservoirs. We describe how to generate
these effective heat reservoirs and detail the strokes of the quantum
Otto cycle. For comparison purposes, we display the efficiency of
the QOHE when the ECI operates under three scenarios: (i) with two
effective thermal reservoirs, (ii) with an effective thermal reservoir
and an effective heat reservoir with a negative apparent temperature,
and (iii) with an effective thermal reservoir and an effective squeezed
thermal reservoir. For all scenarios, we present numerical results
that illustrate the applicability of our scheme by comparing the engine
efficiencies obtained from the physical models with and without the
adiabatic elimination, which is the primary approximation for generating
the effective heat reservoirs mentioned above. Beyond the parameter
regime of this approximation, correlations between the electronic
and motional components of the ion can become significant, complicating
the application of the definitions of heat and work employed in this
study. Nevertheless, if one is interested in exploring such correlations
and other types of reservoirs, the proposed scheme could, in principle,
still be applied. Lastly, we also show how to simulate a QOHE whose
working substance is a quantum harmonic oscillator operating with
the effective heat reservoirs cited above.
\begin{acknowledgments}
We acknowledge financial support from the following Brazilian agencies:
Coordenação de Aperfeiçoamento de Pessoal de Ní­vel Superior (CAPES),
Financial code 001; National Council for Scientific and Technological
Development (CNPq), Grants No. 311612/2021-0 and 301500/2018-5; São
Paulo Research Foundation (FAPESP), Grants No. 2019/11999-5, No. 2021/04672-0,
and No. 2022/10218-2; and Goiás State Research Support Foundation
(FAPEG). This work was performed as part of the Brazilian National
Institute of Science and Technology for Quantum Information (INCT-IQ/CNPq),
Grant No. 465469/2014-0.
\end{acknowledgments}

\onecolumngrid

\appendix

\section{\label{app:A}Effective dynamics of the two-level system}

In this first appendix, we derive the effective master equation presented
in Eq. (\ref{eq:17}) of the main text. To achieve this, we utilize
the projection-based adiabatic elimination method described in Refs.
\citep{Finkelstein-Shapiro2020,Saideh2020} to eliminate the fast
degrees of freedom of the ion.

We begin by writing the Liouvillian superoperator for the ion, considering
the operator-vector isomorphism defined by the map $\left|\psi_{e}\right\rangle \left\langle \psi'_{e}\right|\otimes\left|\psi_{x}\right\rangle \left\langle \psi'_{x}\right|\otimes\left|\psi_{y}\right\rangle \left\langle \psi'_{y}\right|\mapsto\left|\psi_{e}\right\rangle \otimes\left|\psi_{e}\right\rangle ^{*}\otimes\left|\psi_{x}\right\rangle \otimes\left|\psi_{x}\right\rangle ^{*}\otimes\left|\psi_{y}\right\rangle \otimes\left|\psi_{y}\right\rangle ^{*}$.
Thus, the master equation for the ion is mapped into the equation
$\left|\left.\dot{\rho}\left(t\right)\right\rangle \right\rangle ={\cal L}\left|\left.\rho\left(t\right)\right\rangle \right\rangle ,$
where $\left|\left.\rho\left(t\right)\right\rangle \right\rangle $
is the vectorized density operator and ${\cal L}$ is the Liouvillian
superoperator acting on the vector space in which $\left|\left.\rho\left(t\right)\right\rangle \right\rangle $
is defined. By making the tensor products explicit, Eq. (\ref{eq:10})
(with $H_{int}$ given by Eq. (\ref{eq:15})) provides the Liouvillian
superoperator
\begin{equation}
{\cal L}={\cal L}_{e}\otimes I_{2m}+I_{2e}\otimes{\cal L}_{m}+{\cal L}_{int},
\end{equation}
where
\begin{equation}
{\cal L}_{e}=-\frac{i\Delta_{e}}{2}\left(\sigma_{z}\otimes I_{e}-I_{e}\otimes\sigma_{z}\right),\label{eq:A2}
\end{equation}
\begin{equation}
I_{2m}=I_{x}\otimes I_{x}\otimes I_{y}\otimes I_{y},
\end{equation}
\begin{equation}
I_{2e}=I_{e}\otimes I_{e},
\end{equation}
\begin{equation}
{\cal L}_{m}=\frac{\kappa}{2}\left(2a_{x}\otimes a_{x}\otimes I_{2y}-a_{x}^{\dagger}a_{x}\otimes I_{x}\otimes I_{2y}-I_{x}\otimes a_{x}^{\dagger}a_{x}\otimes I_{2y}+2I_{2x}\otimes a_{y}\otimes a_{y}-I_{2x}\otimes a_{y}^{\dagger}a_{y}\otimes I_{y}-I_{2x}\otimes I_{y}\otimes a_{y}^{\dagger}a_{y}\right),\label{eq:A5}
\end{equation}
and
\begin{multline}
{\cal L}_{int}=-i\left(s_{x}\otimes I_{e}\otimes a_{x}^{\dagger}\otimes I_{x}\otimes I_{2y}+s_{x}^{\dagger}\otimes I_{e}\otimes a_{x}\otimes I_{x}\otimes I_{2y}-I_{e}\otimes s_{x}\otimes I_{x}\otimes a_{x}^{\dagger}\otimes I_{2y}-I_{e}\otimes s_{x}^{\dagger}\otimes I_{x}\otimes a_{x}\otimes I_{2y}\right)\\
-i\left(s_{y}\otimes I_{e}\otimes I_{2x}\otimes a_{y}^{\dagger}\otimes I_{y}+s_{y}^{\dagger}\otimes I_{e}\otimes I_{2x}\otimes a_{y}\otimes I_{y}-I_{e}\otimes s_{y}\otimes I_{2x}\otimes I_{y}\otimes a_{y}^{\dagger}-I_{e}\otimes s_{y}^{\dagger}\otimes I_{2x}\otimes I_{y}\otimes a_{y}\right).\label{eq:A6}
\end{multline}
Here, $I_{e}$ and $I_{x}\otimes I_{y}$ are the identity operators
of the electronic and motional parts of the ion, respectively.

As described in Sec. \ref{sec:IV}, we require that $\kappa\gg\lambda\Omega_{\alpha,i}$
($\alpha=x,y$ and $i=1,2$), which ensures that the motional component
of the ion (MCI) quickly reaches the steady state $\left|\left.\rho_{m}^{ss}\right\rangle \right\rangle =\left|0_{x}\right\rangle \otimes\left|0_{x}\right\rangle \otimes\left|0_{y}\right\rangle \otimes\left|0_{y}\right\rangle $.
In this regime, the motional dynamics approximately occur within the
subspace spanned by the basis presented in Tab. \ref{tab:1}. Thus,
the slow dynamics of the MCI takes place in the subspace described
by basis $\left\{ \left|\left.\psi_{0}\right\rangle \right\rangle =\left|\left.\rho_{m}^{ss}\right\rangle \right\rangle \right\} $,
while the fast dynamics of the MCI (which we aim to eliminate) occurs
in the subspace generated by basis $\left\{ \left|\left.\psi_{1}\right\rangle \right\rangle ,\left|\left.\psi_{2}\right\rangle \right\rangle ,\ldots\left|\left.\psi_{15}\right\rangle \right\rangle \right\} $.
\begin{center}
\begin{table}[H]
\begin{centering}
\begin{tabular}{|c|c|c|c|}
\hline 
$\left|\left.\psi_{0}\right\rangle \right\rangle =\left|0_{x}\right\rangle \otimes\left|0_{x}\right\rangle \otimes\left|0_{y}\right\rangle \otimes\left|0_{y}\right\rangle $ & $\left|\left.\psi_{1}\right\rangle \right\rangle =\left|0_{x}\right\rangle \otimes\left|0_{x}\right\rangle \otimes\left|1_{y}\right\rangle \otimes\left|0_{y}\right\rangle $ & $\left|\left.\psi_{2}\right\rangle \right\rangle =\left|0_{x}\right\rangle \otimes\left|0_{x}\right\rangle \otimes\left|0_{y}\right\rangle \otimes\left|1_{y}\right\rangle $ & $\left|\left.\psi_{3}\right\rangle \right\rangle =\left|0_{x}\right\rangle \otimes\left|0_{x}\right\rangle \otimes\left|1_{y}\right\rangle \otimes\left|1_{y}\right\rangle $\tabularnewline
\hline 
$\left|\left.\psi_{4}\right\rangle \right\rangle =\left|1_{x}\right\rangle \otimes\left|0_{x}\right\rangle \otimes\left|0_{y}\right\rangle \otimes\left|0_{y}\right\rangle $ & $\left|\left.\psi_{5}\right\rangle \right\rangle =\left|1_{x}\right\rangle \otimes\left|0_{x}\right\rangle \otimes\left|1_{y}\right\rangle \otimes\left|0_{y}\right\rangle $ & $\left|\left.\psi_{6}\right\rangle \right\rangle =\left|1_{x}\right\rangle \otimes\left|0_{x}\right\rangle \otimes\left|0_{y}\right\rangle \otimes\left|1_{y}\right\rangle $ & $\left|\left.\psi_{7}\right\rangle \right\rangle =\left|1_{x}\right\rangle \otimes\left|0_{x}\right\rangle \otimes\left|1_{y}\right\rangle \otimes\left|1_{y}\right\rangle $\tabularnewline
\hline 
$\left|\left.\psi_{8}\right\rangle \right\rangle =\left|0_{x}\right\rangle \otimes\left|1_{x}\right\rangle \otimes\left|0_{y}\right\rangle \otimes\left|0_{y}\right\rangle $ & $\left|\left.\psi_{9}\right\rangle \right\rangle =\left|0_{x}\right\rangle \otimes\left|1_{x}\right\rangle \otimes\left|1_{y}\right\rangle \otimes\left|0_{y}\right\rangle $ & $\left|\left.\psi_{10}\right\rangle \right\rangle =\left|0_{x}\right\rangle \otimes\left|1_{x}\right\rangle \otimes\left|0_{y}\right\rangle \otimes\left|1_{y}\right\rangle $ & $\left|\left.\psi_{11}\right\rangle \right\rangle =\left|0_{x}\right\rangle \otimes\left|1_{x}\right\rangle \otimes\left|1_{y}\right\rangle \otimes\left|1_{y}\right\rangle $\tabularnewline
\hline 
$\left|\left.\psi_{12}\right\rangle \right\rangle =\left|1_{x}\right\rangle \otimes\left|1_{x}\right\rangle \otimes\left|0_{y}\right\rangle \otimes\left|0_{y}\right\rangle $ & $\left|\left.\psi_{13}\right\rangle \right\rangle =\left|1_{x}\right\rangle \otimes\left|1_{x}\right\rangle \otimes\left|1_{y}\right\rangle \otimes\left|0_{y}\right\rangle $ & $\left|\left.\psi_{14}\right\rangle \right\rangle =\left|1_{x}\right\rangle \otimes\left|1_{x}\right\rangle \otimes\left|0_{y}\right\rangle \otimes\left|1_{y}\right\rangle $ & $\left|\left.\psi_{15}\right\rangle \right\rangle =\left|1_{x}\right\rangle \otimes\left|1_{x}\right\rangle \otimes\left|1_{y}\right\rangle \otimes\left|1_{y}\right\rangle $\tabularnewline
\hline 
\end{tabular}
\par\end{centering}
\caption{\label{tab:1}Truncated basis employed to describe the motional component
of the ion.}
\end{table}
\par\end{center}

Following the projection-based adiabatic elimination method presented
in Refs. \citep{Finkelstein-Shapiro2020,Saideh2020}, the effective
Liouvillian superoperator of the ion is given by (see Eq. (19) of
Ref. \citep{Saideh2020}) 
\begin{equation}
{\cal L}_{eff}={\cal L}_{e}\otimes{\cal P}_{m}+{\cal P}{\cal L}_{int}{\cal P}-{\cal P}{\cal L}_{int}{\cal Q}\left[I_{2e}\otimes\left({\cal Q}_{m}{\cal L}_{m}{\cal Q}_{m}\right)^{-1}\right]\left(I_{2e}\otimes{\cal Q}_{m}{\cal L}_{m}{\cal P}_{m}+{\cal Q}{\cal L}_{int}{\cal P}\right),\label{eq:A7}
\end{equation}
in which ${\cal P}=I_{2e}\otimes{\cal P}_{m}$ and ${\cal Q}=I_{2e}\otimes{\cal Q}_{m},$with
${\cal P}_{m}=\left|\left.\rho_{m}^{ss}\right\rangle \right\rangle \left\langle \left\langle I_{m}\right.\right|$
($I_{m}=I_{x}\otimes I_{y}$) and $Q_{m}=I_{2m}-{\cal P}_{m}$. Note
that the projector ${\cal P}$ projects the ion dynamics onto the
slow subspace, since ${\cal P}\left|\left.\rho\left(t\right)\right\rangle \right\rangle =\left|\left.\rho_{e}\left(t\right)\right\rangle \right\rangle \otimes\left|\left.\rho_{m}^{ss}\right\rangle \right\rangle $,
while the projector ${\cal Q}$ projects the ion dynamics onto the
fast subspace. By taking into account Eqs. (\ref{eq:A2}), (\ref{eq:A5})
and (\ref{eq:A6}), we have that
\begin{equation}
{\cal P}{\cal L}_{int}{\cal P}=0,\label{eq:A9}
\end{equation}
\begin{multline}
{\cal P}{\cal L}_{int}{\cal Q}=-i\left[\left(s_{x}^{\dagger}\otimes I_{e}-I_{e}\otimes s_{x}\right)\otimes\left|\left.\rho_{m}^{ss}\right\rangle \right\rangle \left(\left\langle \left\langle \psi_{4}\right.\right|+\left\langle \left\langle \psi_{7}\right.\right|\right)+\left(s_{x}\otimes I_{e}-I_{e}\otimes s_{x}^{\dagger}\right)\otimes\left|\left.\rho_{m}^{ss}\right\rangle \right\rangle \left(\left\langle \left\langle \psi_{8}\right.\right|+\left\langle \left\langle \psi_{11}\right.\right|\right)\right.\\
\left.+\left(s_{y}^{\dagger}\otimes I_{e}-I_{e}\otimes s_{y}\right)\otimes\left|\left.\rho_{m}^{ss}\right\rangle \right\rangle \left(\left\langle \left\langle \psi_{1}\right.\right|+\left\langle \left\langle \psi_{13}\right.\right|\right)+\left(s_{y}\otimes I_{e}-I_{e}\otimes s_{y}^{\dagger}\right)\otimes\left|\left.\rho_{m}^{ss}\right\rangle \right\rangle \left(\left\langle \left\langle \psi_{2}\right.\right|+\left\langle \left\langle \psi_{14}\right.\right|\right)\right],
\end{multline}
\begin{equation}
{\cal Q}_{m}{\cal L}_{m}{\cal P}_{m}=0,
\end{equation}
and
\begin{equation}
{\cal Q}{\cal L}_{int}{\cal P}=-i\left(s_{x}\otimes I_{e}\otimes\left|\left.\psi_{4}\right\rangle \right\rangle \left\langle \left\langle I_{m}\right.\right|-I_{e}\otimes s_{x}\otimes\left|\left.\psi_{8}\right\rangle \right\rangle \left\langle \left\langle I_{m}\right.\right|+s_{y}\otimes I_{e}\otimes\left|\left.\psi_{1}\right\rangle \right\rangle \left\langle \left\langle I_{m}\right.\right|-I_{e}\otimes s_{y}\otimes\left|\left.\psi_{2}\right\rangle \right\rangle \left\langle \left\langle I_{m}\right.\right|\right).
\end{equation}
Furthermore, with the help of Sympy, we have 
\begin{multline}
\left({\cal Q}_{m}{\cal L}_{m}{\cal Q}_{m}\right)^{-1}=-\frac{1}{\kappa}\left(2\left|\left.\psi_{1}\right\rangle \right\rangle \left\langle \left\langle \psi_{1}\right.\right|+2\left|\left.\psi_{2}\right\rangle \right\rangle \left\langle \left\langle \psi_{2}\right.\right|+\left|\left.\psi_{3}\right\rangle \right\rangle \left\langle \left\langle \psi_{3}\right.\right|+2\left|\left.\psi_{4}\right\rangle \right\rangle \left\langle \left\langle \psi_{4}\right.\right|+\left|\left.\psi_{5}\right\rangle \right\rangle \left\langle \left\langle \psi_{5}\right.\right|\right.\\
+\left|\left.\psi_{6}\right\rangle \right\rangle \left\langle \left\langle \psi_{6}\right.\right|+\frac{2}{3}\left|\left.\psi_{7}\right\rangle \right\rangle \left\langle \left\langle \psi_{7}\right.\right|+2\left|\left.\psi_{8}\right\rangle \right\rangle \left\langle \left\langle \psi_{8}\right.\right|+\left|\left.\psi_{9}\right\rangle \right\rangle \left\langle \left\langle \psi_{9}\right.\right|+\left|\left.\psi_{10}\right\rangle \right\rangle \left\langle \left\langle \psi_{10}\right.\right|+\frac{2}{3}\left|\left.\psi_{11}\right\rangle \right\rangle \left\langle \left\langle \psi_{11}\right.\right|\\
+\left|\left.\psi_{12}\right\rangle \right\rangle \left\langle \left\langle \psi_{12}\right.\right|+\frac{2}{3}\left|\left.\psi_{13}\right\rangle \right\rangle \left\langle \left\langle \psi_{13}\right.\right|+\frac{2}{3}\left|\left.\psi_{14}\right\rangle \right\rangle \left\langle \left\langle \psi_{14}\right.\right|+\frac{1}{2}\left|\left.\psi_{15}\right\rangle \right\rangle \left\langle \left\langle \psi_{15}\right.\right|+\frac{4}{3}\left|\left.\psi_{4}\right\rangle \right\rangle \left\langle \left\langle \psi_{7}\right.\right|\\
\left.+\frac{4}{3}\left|\left.\psi_{8}\right\rangle \right\rangle \left\langle \left\langle \psi_{11}\right.\right|+\frac{1}{2}\left|\left.\psi_{12}\right\rangle \right\rangle \left\langle \left\langle \psi_{15}\right.\right|+\frac{4}{3}\left|\left.\psi_{1}\right\rangle \right\rangle \left\langle \left\langle \psi_{13}\right.\right|+\frac{4}{3}\left|\left.\psi_{2}\right\rangle \right\rangle \left\langle \left\langle \psi_{14}\right.\right|+\frac{1}{2}\left|\left.\psi_{3}\right\rangle \right\rangle \left\langle \left\langle \psi_{15}\right.\right|\right).\label{eq:A12}
\end{multline}
Finally, substituting Eq. (\ref{eq:A9})-(\ref{eq:A12}) into Eq.
(\ref{eq:A7}), we obtain
\begin{equation}
{\cal L}_{eff}=\left[{\cal L}_{e}+\sum_{\alpha=x,y}\frac{2}{\kappa}\left(2s_{\alpha}\otimes s_{\alpha}-s_{\alpha}^{\dagger}s_{\alpha}\otimes I_{e}-I_{e}\otimes s_{\alpha}^{\dagger}s_{\alpha}\right)\right]\otimes{\cal P}_{m},
\end{equation}
which leads to Eq. (\ref{eq:17}).

\section{\label{app:B}Effective dynamics of the quantum harmonic oscillator}

Here, in this second appendix, we obtain the effective master equation
for the motional state shown in Eq. (\ref{eq:54}) of the main text.
As in the first appendix, here we also follow the projection-based
adiabatic elimination method introduced in Refs. \citep{Finkelstein-Shapiro2020,Saideh2020}.

Considering the operator-vector isomorphism mentioned in the previous
appendix, Eq. (\ref{eq:48}) (with $H_{int}$ defined by Eq. (\ref{eq:52}))
leads to the Liouvillian superoperator
\begin{equation}
{\cal L}={\cal L}_{e}\otimes I_{2m}+I_{2e}\otimes{\cal L}_{m}+{\cal L}_{int},
\end{equation}
where
\begin{equation}
{\cal L}_{e}=\frac{\gamma_{ge}}{2}\left(2\sigma_{ge}\otimes\sigma_{ge}-\sigma_{ge}^{\dagger}\sigma_{ge}\otimes I_{e}-I_{e}\otimes\sigma_{ge}^{\dagger}\sigma_{ge}\right)+\frac{\gamma_{gf}}{2}\left(2\sigma_{gf}\otimes\sigma_{gf}-\sigma_{gf}^{\dagger}\sigma_{gf}\otimes I_{e}-I_{e}\otimes\sigma_{gf}^{\dagger}\sigma_{gf}\right),\label{eq:B2}
\end{equation}
\begin{equation}
I_{2m}=I_{m}\otimes I_{m},
\end{equation}
\begin{equation}
{\cal L}_{m}=-i\Delta_{m}\left(a^{\dagger}a\otimes I_{m}-I_{m}\otimes a^{\dagger}a\right),\label{eq:B4}
\end{equation}
and
\begin{multline}
{\cal L}_{int}=-i\left(\sigma_{ge}\otimes I_{e}\otimes s_{ge}^{\dagger}\otimes I_{m}+\sigma_{ge}^{\dagger}\otimes I_{e}\otimes s_{ge}\otimes I_{m}-I_{e}\otimes\sigma_{ge}\otimes I_{m}\otimes s_{ge}^{\dagger}-I_{e}\otimes\sigma_{ge}^{\dagger}\otimes I_{m}\otimes s_{ge}\right)\\
-i\left(\sigma_{gf}\otimes I_{e}\otimes s_{gf}^{\dagger}\otimes I_{m}+\sigma_{gf}^{\dagger}\otimes I_{e}\otimes s_{gf}\otimes I_{m}-I_{e}\otimes\sigma_{gf}\otimes I_{m}\otimes s_{gf}^{\dagger}-I_{e}\otimes\sigma_{gf}^{\dagger}\otimes I_{m}\otimes s_{gf}\right).\label{eq:B5}
\end{multline}

As stated in Sec. (\ref{sec:VII}), we assume that $\gamma_{\alpha}\gg\lambda\Omega_{\alpha,i}$
($\alpha=ge,ef$ and $i=1,2$), which implies that the ECI rapidly
reaches the steady state $\left|\left.\rho_{e}^{ss}\right\rangle \right\rangle =\left|g\right\rangle \otimes\left|g\right\rangle $.
With this, the slow and fast subspaces of the ECI is spanned by the
basis $\left\{ \left|\left.\rho_{e}^{ss}\right\rangle \right\rangle =\left|g\right\rangle \otimes\left|g\right\rangle \right\} $
and $\left\{ \left|g\right\rangle \otimes\left|e\right\rangle ,\left|g\right\rangle \otimes\left|f\right\rangle ,\left|e\right\rangle \otimes\left|g\right\rangle ,\left|e\right\rangle \otimes\left|e\right\rangle ,\left|e\right\rangle \otimes\left|f\right\rangle ,\left|f\right\rangle \otimes\left|g\right\rangle ,\left|f\right\rangle \otimes\left|e\right\rangle ,\left|f\right\rangle \otimes\left|f\right\rangle \right\} $,
respectively.

According to Ref. \citep{Saideh2020}, the effective Liouvillian superoperator
for the ion has the form
\begin{equation}
{\cal L}_{eff}={\cal P}_{e}\otimes{\cal L}_{m}+{\cal P}{\cal L}_{int}{\cal P}-{\cal P}{\cal L}_{int}{\cal Q}\left[\left({\cal Q}_{e}{\cal L}_{e}{\cal Q}_{e}\right)^{-1}\otimes I_{2m}\right]\left({\cal Q}_{e}{\cal L}_{e}{\cal P}_{e}\otimes I_{2m}+{\cal Q}{\cal L}_{int}{\cal P}\right),\label{eq:B6}
\end{equation}
where ${\cal P}={\cal P}_{e}\otimes I_{2m}$ and ${\cal Q}={\cal Q}_{e}\otimes I_{2m}$,
with ${\cal P}_{e}=\left|\left.\rho_{e}^{ss}\right\rangle \right\rangle \left\langle \left\langle I_{e}\right.\right|$
and ${\cal Q}_{e}=I_{2e}-{\cal P}_{e}$. By applying these projectors
in Eqs. (\ref{eq:A2}), (\ref{eq:A5}) and (\ref{eq:A6}), we obtain
\begin{equation}
{\cal P}{\cal L}_{int}{\cal P}=0,\label{eq:B7}
\end{equation}
\begin{multline}
{\cal P}{\cal L}_{int}{\cal Q}=-i\left[\left|\left.\rho_{e}^{ss}\right\rangle \right\rangle \left\langle \psi_{3}\right|\otimes\left(s_{ge}^{\dagger}\otimes I_{m}-I_{m}\otimes s_{ge}\right)+\left|\left.\rho_{e}^{ss}\right\rangle \right\rangle \left\langle \psi_{1}\right|\otimes\left(s_{ge}\otimes I_{m}-I_{m}\otimes s_{ge}^{\dagger}\right)\right.\\
\left.+\left|\left.\rho_{e}^{ss}\right\rangle \right\rangle \left\langle \psi_{6}\right|\otimes\left(s_{gf}^{\dagger}\otimes I_{m}-I_{m}\otimes s_{gf}\right)+\left|\left.\rho_{e}^{ss}\right\rangle \right\rangle \left\langle \psi_{2}\right|\otimes\left(s_{gf}\otimes I_{m}-I_{m}\otimes s_{gf}^{\dagger}\right)\right],
\end{multline}
\begin{equation}
{\cal Q}_{e}{\cal L}_{e}{\cal P}_{e}=0,
\end{equation}
and
\begin{equation}
{\cal Q}{\cal L}_{int}{\cal P}=-i\left(\left|\psi_{3}\right\rangle \left\langle \left\langle I_{e}\right.\right|\otimes s_{ge}\otimes I_{m}-\left|\psi_{1}\right\rangle \left\langle \left\langle I_{e}\right.\right|\otimes I_{m}\otimes s_{ge}+\left|\psi_{6}\right\rangle \left\langle \left\langle I_{e}\right.\right|\otimes s_{gf}\otimes I_{m}-\left|\psi_{2}\right\rangle \left\langle \left\langle I_{e}\right.\right|\otimes I_{m}\otimes s_{gf}\right).
\end{equation}
Furthermore, using Sympy, we have that
\begin{multline}
\left({\cal Q}_{e}{\cal L}_{e}{\cal Q}_{e}\right)^{-1}=-\frac{2}{\gamma_{ge}}\left|\left.\psi_{1}\right\rangle \right\rangle \left\langle \left\langle \psi_{1}\right.\right|-\frac{2}{\gamma_{gf}}\left|\left.\psi_{2}\right\rangle \right\rangle \left\langle \left\langle \psi_{2}\right.\right|-\frac{2}{\gamma_{ge}}\left|\left.\psi_{3}\right\rangle \right\rangle \left\langle \left\langle \psi_{3}\right.\right|-\frac{1}{\gamma_{ge}}\left|\left.\psi_{4}\right\rangle \right\rangle \left\langle \left\langle \psi_{4}\right.\right|\\
-\frac{2}{\gamma_{ge}+\gamma_{gf}}\left|\left.\psi_{5}\right\rangle \right\rangle \left\langle \left\langle \psi_{5}\right.\right|-\frac{2}{\gamma_{gf}}\left|\left.\psi_{6}\right\rangle \right\rangle \left\langle \left\langle \psi_{6}\right.\right|-\frac{2}{\gamma_{ge}+\gamma_{gf}}\left|\left.\psi_{7}\right\rangle \right\rangle \left\langle \left\langle \psi_{7}\right.\right|-\frac{1}{\gamma_{gf}}\left|\left.\psi_{8}\right\rangle \right\rangle \left\langle \left\langle \psi_{8}\right.\right|.\label{eq:B11}
\end{multline}
Lastly, by replacing Eqs. (\ref{eq:B7})-(\ref{eq:B11}) into Eq.
(\ref{eq:B6}), we reach
\begin{equation}
{\cal L}_{eff}={\cal P}_{e}\otimes\left[{\cal L}_{m}+\sum_{\alpha=ge,gf}\frac{2}{\gamma_{\alpha}}\left(2s_{\alpha}\otimes s_{\alpha}-s_{\alpha}^{\dagger}s_{\alpha}\otimes I_{e}-I_{e}\otimes s_{\alpha}^{\dagger}s_{\alpha}\right)\right],
\end{equation}
leading to Eq. (\ref{eq:54}).

\twocolumngrid

\bibliographystyle{apsrev4-2}
\bibliography{References}

\end{document}